\documentstyle[12pt]{article}

\def\noi{\noindent}
\def\ad{\rm ad}

\def\DD{\hbox{{\rm \bf D}}}
\def\ket{\rangle}
\def\bra{\langle}
\def\be{\begin{equation}}
\def\bea{\begin{eqnarray}}
\def\nn{\nonumber}
\def\ee{\end{equation}}
\def\eea{\end{eqnarray}}
\def\CA{\hbox{{$\cal A$}}}

\def\CR{\hbox{{$\cal R$}}}

\def\one{1\!\!1}
\def\INT{{\textstyle \int\kern-.642em\int}}
\def\R{{R\kern-.647em R}}
\def\C{{C\kern-.647em C}}
\def\T{{T\kern-.647em T}}
\def\Q{{Q\kern-.647em Q}}
\def\F{{F\kern-.647em F}}
\def\Z{{Z\kern-.647em Z}}
\def\N{{N\kern-.79em N}}
\def\lform{\hbox{$\sqcup$}\llap{\hbox{$\sqcap$}}}

\def\defe{\stackrel{{\rm def}}{=}}

\def\ad{{\rm ad}}

\def\proof{\goodbreak\noindent{\bf Proof\quad}}
\def\endproof{{\ $\lform$}\bigskip }

\begin{document}
\begin{titlepage}
\rightline{PRA-HEP-93/2}
\vskip 2em
\begin{center}
{\large \bf THE QUANTUM-DOUBLE FOR A NONSTANDARD DEFORMATION}\\
{\large \bf OF A BOREL SUBALGEBRA $sl(2,C)$}\\[6em]
 \v{C} Burd\'{\i}k${}^{\diamond }$\footnote{E-mail:
Bitnet=(VSETIN AT CSPUNI12)},
and P. Hellinger${}^{*}$\footnote{E-mail:
Bitnet=(HELINGER AT CSPUNI12)} \\[2em]
{\sl ${}^{*}$Department of Theoretical Physics and
 ${}^{\diamond}$Nuclear Centre,\\
Faculty of Mathematics and Physics,Charles University,\\
V Ho\-le\-\v{s}o\-vi\v{c}k\'{a}ch 2, 180 00 Prague 8, Czech Republic}
\end{center}
\vskip 3cm

\begin{abstract}
 We give a construction of Drienfeld's quantum double
 for a nonstandard deformation of Borel subalgebra of $sl(2)$.
 We construct explicitly some simple representations of this
 quantum algebra and from the universal R-matrix we obtain the explicit
 solutions of the Yang-Baxter equation in those cases.
\end{abstract}

\vskip 4.5cm

\noi PRA-HEP-93/2
\noi hep-th/9303035
\noi March 1993

\end{titlepage}

\setcounter{page}{1}

\section{ Introduction}
\vspace{10pt}

One of the most important quantum group constructions is Drienfeld's
quantum double \cite{Dri:qg87}. This construction was
 used by Drienfeld for construction
of quantum deformations of simple Lie algebras. In well know example of
$U(sl(2))$ we start from a Borel subalgebra and the Drienfeld's quantum
double gives $U_q(sl(2)) \otimes U(u(1))$.

It was pointed by Ogievetski \cite{Ogi:ta93} that there exists
another nonequivalent
Hopf algebra deformation of the Borel subalgebra
connected via FRT construction~\cite{FRT}
with the solution of the Yang-Baxter equation
\be
R = \pmatrix{
     1& -\gamma&\gamma& -\gamma\beta \cr
   0& 1& 0& \beta \cr
  0 & 0& 1& -\beta \cr
   0 &0 &0&1 }
\label{man}
\ee
which was found in \cite{DMMZ} and studied in \cite{EOW1}
\cite{EOW2} \cite{Hie:92} \cite{Hla:92} \cite{Zak:hs91}
The aim of this letter
is solve an interesting question what we will obtain
when the Drienfeld's quantum double construction
 is applied to this deformation.

\section{The basic Hopf algebra and its dual}
\vspace{10pt}

Let us consider the Hopf algebra {\CA}=$U_{\gamma}b_{-}$
 over \C[[$\gamma$]],
the formal deformation of $U b_{-}$ of $sl(2)$ with the
generators $\tau,\ \pi$ and with the relations
\begin{equation}
[\tau,\ \pi]\ =\ -2\pi.
\label{com}
\end{equation}
The coalgebra structure is given by
\begin{eqnarray}
\Delta (\tau)&=& \tau\otimes 1 + \Lambda\otimes\tau \nn\\
\Delta (\pi)&=& \pi \otimes 1+ \Lambda^{-1}\otimes\pi\nn\\
\epsilon (\tau)&=& \epsilon (\pi)\ =\ 0
\end{eqnarray}
and antipodes are
\begin{eqnarray}
S(\tau)&=& - \Lambda^{-1} \tau\nn\\
S(\pi)&=&- \Lambda \pi\nn\\
S^{\sigma}(\tau)&=& -\tau \Lambda^{-1}\nn\\
S^{\sigma}(\pi)&=&- \Lambda \pi,
\end{eqnarray}
where $\Lambda=(1- \gamma \pi)^{-1}$.

We may construct the dual Hopf algebra {$\CA^*$}
 with the generators
$ T,\ P$ and with the pairing
\begin{eqnarray}
\bra T,\ \tau^{n}\pi^{k}{\ket} &\defe& \delta_{n1} \delta_{k0}
\label{deft}\\
{\bra} P,\ \tau^{n}\pi^{k}{\ket} &\defe& \delta_{n0} \delta_{k1}
\label{defp}
\end{eqnarray}
{}From this pairing one may easily deduce the comultiplications
\begin{eqnarray}
\Delta(T) &=& T\otimes 1 + 1\otimes T\nn\\
\Delta(P) &=& P\otimes e^{2T} + 1\otimes P
\end{eqnarray}
\proof
\begin{eqnarray*}
 {\bra} \Delta( T),\ \tau^{n_1}\pi^{k_1}
\otimes\tau^{n_2}\pi^{k_2}{\ket} &=&
{\bra} T,\  \tau^{n_1}\pi^{k_1}
\tau^{n_2}\pi^{k_2}{\ket} \ = \\ &=&
\ \delta_{n_{1}1} \delta_{k_{1}0} \delta_{n_{2}0} \delta_{k_{2}0}+
 \delta_{n_{1}0} \delta_{k_{1}0} \delta_{n_{2}1} \delta_{k_{2}0}
\end{eqnarray*}
and
\begin{eqnarray*}{\bra} \Delta( P),\ \tau^{n_1}\pi^{k_1}
\otimes\tau^{n_2}\pi^{k_2}{\ket} &=&
{\bra} P,\  \tau^{n_1}\pi^{k_1}
\tau^{n_2}\pi^{k_2}{\ket} \ = \\&=&
 \delta_{n_{1}0} \delta_{k_{1}0} \delta_{n_{2}0} \delta_{k_{2}1}+
 \delta_{n_{1}0} \delta_{k_{1}1} \delta_{k_{2}0} 2^{n_2}
\end{eqnarray*}
and
$${\bra} T^{k},\ \tau^{n}{\ket} \ =\ n!\delta_{nk}$$ as will be proven
(see the equation~(\ref{tn})).
\endproof

We are to find the commutation relations between $T$ and $P$.
The result is
\begin{equation}
[T,\ P]\ =\ - {1\over2} \gamma( e^{2T} -1)
\end{equation}
\proof
\begin{eqnarray*}
{\bra} TP,\ \tau^n\pi^k{\ket} &=& {\bra} T\otimes P,\ \Delta(\tau^n\pi^k){\ket}
\ =\\
&=&\delta_{n1}\delta_{k1}\\
{\bra} PT,\ \tau^n\pi^k{\ket} &=&{\bra} P\otimes T,\ \Delta(\tau^n\pi^k){\ket}
\
 =\\
&=&\delta_{n1} \delta_{k1}+ (n > 0) \gamma \delta_{k0} 2^{n-1},
\end{eqnarray*}
where the symbol $(x)$ is defined
\[ (x) = \left\{
\begin{array}{ll}
1& {{\textstyle{\rm if\ \it x}\ is\ true}}\\
0&{\textstyle\rm otherwise}
\end{array}
\right. \]
\endproof

Now is time for counting of powers of the generators.
Let us start with promised
\begin{equation}
{\bra}  T^{n},\ \tau^k\pi^m{\ket} \ =\ n!\delta_{nk}\delta_{m0}
\label{tn}
\end{equation}
\proof
\begin{eqnarray*}
{\bra}  T^{n},\ \tau^k\pi^m{\ket} &=& {\bra}  T^{\otimes n},\ \Delta^{n-1}
(\tau^k\pi^m){\ket} \ =\\
&=&\delta_{m0}{\bra}  T^{\otimes n},\ (t_1+\cdots+t_n)^{k}{\ket}
\end{eqnarray*}
since (\ref{deft}), (\ref{com}) and
$\Lambda=1+\gamma \pi + o(\gamma^2)$, where
\begin{eqnarray*}
t_1&\defe& t\otimes 1\otimes \cdots\otimes 1\\
t_2&\defe& 1\otimes t\otimes 1\otimes \cdots\otimes 1\\
&\vdots&\\
t_{n-1}&\defe& 1\otimes \cdots\otimes 1 \otimes t \otimes 1\\
t_n&\defe& 1\otimes \cdots\otimes 1\otimes t
\end{eqnarray*}
And so $
{\bra}  T^{\otimes n},\ (t_1+\cdots+t_n)^{k}{\ket} \ =\
\delta_{nk} n!$\endproof

And now we will switch to $P$
\begin{equation}
{\bra} P^n,\ \tau^m\pi^k{\ket} \ =\ \delta_{m0} (-\gamma)^{n-k}
\Theta^n_k
\end{equation}
where
$$ \Theta^n_k \ =\ \sum_{p=0}^{k}{k\choose p}(-1)^{k-p} p^n$$
with the following properties
\begin{eqnarray}
\Theta^n_k\ =\ 0\ {\textstyle\rm for}\ k > n\quad \quad
\Theta^n_n \ =\ n!
\end{eqnarray}
\proof
\begin{eqnarray*}
\Delta(\pi)&=& \pi\otimes 1  +1\otimes\pi -\gamma
\pi\otimes\pi\\
\Delta^{n}(\pi)&=& \sum_{i=1}^{n+1} \pi_{i}+(-\gamma)
\sum_{i> j=1}^{n+1}\pi_{i}\pi_{j}+(-\gamma)^2
\sum_{i> j>k=1}^{n+1}\pi_{i}\pi_{j}\pi_{k}+\cdots+\\
& &+
(-\gamma)^{n}\pi_{1}\pi_{2}\cdots\pi_{n}\pi_{n+1},
\end{eqnarray*}
by the induction, where $\pi_{i}$ is defined similar as
for $\tau$. We ought to calculate
\begin{eqnarray*}
{\bra}  P^{n},\ \tau^k\pi^m{\ket} &=& {\bra}  P^{\otimes n},\ \Delta^{n-1}
(\tau^m\pi^k){\ket} \ =\\
&=& \delta_{m0}{\bra}  P^{\otimes n},\ \Delta^{n-1}
(\pi)^k{\ket}
\end{eqnarray*}
Thanks to (\ref{defp}) we may $\pi_{j}$'s in the previous
pairing consider as grassman\-ian variables $\pi_{j}^2=0$,
for $j=1,\ \ldots,\ n$ and $\pi_{i}\pi_{j}=
\pi_{j}\pi_{i}$ for $i,\ j=1,\ \ldots,\ n$.
Under such ansatz we have this very convenient result
\begin{equation}
\Delta^{n}(\pi)\ \stackrel{{\bra} ,{\ket} }{=}\ {-{1\over\gamma}}
(e^{-\gamma \sum_{i=1}^{n+1}\pi_{1}}-1)
\end{equation}
as one may easily check ($\stackrel{{\bra} ,{\ket} }{=}$ means the
equality in the pairing). So we will proceed
\begin{eqnarray*}
{\bra}  P^{\otimes n},\ \Delta^{n-1}
(\pi)^k{\ket} &=&{\bra}  P^{\otimes n},\ ({-{1\over\gamma}}
(e^{-\gamma \sum_{i=1}^{n}\pi_{1}}-1))^k{\ket} \ =\\
&=&{\bra}  P^{\otimes n},\ (-\gamma)^{-k}\sum_{p=0}^{k}
{k\choose p}(-1)^{k-p} e^{-\gamma p \sum_{i=1}^{n}\pi_{i}}{\ket}
\ =\\ &=&
(-\gamma)^{n-k}
\sum_{p=0}^{k}{k\choose p}(-1)^{k-p} p^n
\end{eqnarray*}
\endproof

Now we may obtain the general formula
\be
{\bra}T^j P^n,\ \tau^m\pi^k{\ket} \ =\ m!\delta_{mj} (-\gamma)^{n-k}
\sum_{p=0}^{k}{k\choose p}(-1)^{k-p} p^n
\label{pair}
\ee
\proof Straightforward
\endproof

At this moment we have the description of the dual Hopf
algebra $\CA^*$ or precisely we have only the explicit
form of the bialgebra structure, but henceforth
we will not need the antipodes. Let us construct the
double now.

\section{The quantum double \protect \DD(\CA)}
\vspace{10pt}

We have to combine $\CA$ and $\CA^{\circ}$, where
$\CA^{\circ}$ is the dual $\CA^{*}$ with the opposite
commultiplication. Let us briefly summarize these
structures:
\begin{eqnarray}
[\tau,\ \pi]&=& -2\pi
\nn \\
{ [T,\;  P ]}  &=& -{1\over 2} \gamma( e^{2T} -1)\nn\\
\Delta (\tau)&=& \tau\otimes 1 + \Lambda\otimes\tau\nn\\
\Delta (\pi)&=& \pi \otimes 1+ \Lambda^{-1}\otimes\pi\nn\\
\Delta(T) &=& T\otimes 1 + 1\otimes T\nn\\
\Delta(P) &=& P\otimes 1+e^{2T}\otimes P
\end{eqnarray}
For the commutation relations between $\pi,\ \tau$ and
$P,\ T$ we are forced to use the standard procedure
\be
X\zeta= \Sigma(\bra \Delta^2(X),\
 S^{\sigma}_{1}
 \Delta^2(\zeta)\ket_{13})
\ee
for $X\in \{ P,\ T\}$ and $\zeta \in \{\pi,\ \tau\}$, where
$\bra,\ket_{13}$ is the evaluation
between the dual algebras on the first and the third
position and $\Sigma$ exchanges $\CA^{\circ}$ and $\CA$

This reads in our case as follows:
\bea
{[P,\ \tau]}&=& -\gamma\tau-2P\nn\\
{[P,\ \pi]}&=& \Lambda^{-1}e^{2T} -1+\gamma \pi\nn\\
{[T,\ \tau]}&=& \Lambda-1\nn\\
{[T,\ \pi]}&=& 0
\eea
\proof
\begin{eqnarray*}
P\tau&=&\Sigma(\bra
e^{2T}\otimes e^{2T}\otimes P+
e^{2T}\otimes P\otimes 1+
P\otimes 1\otimes 1
,\\ & & S^{\sigma}_{1}(
\tau\otimes 1\otimes 1+
\Lambda\otimes \tau\otimes 1+
\Lambda\otimes \Lambda\otimes \tau)
\ket_{13})\ =\\
&=&
\Sigma(\bra
e^{2T}\otimes e^{2T}\otimes P+
e^{2T}\otimes P\otimes 1+
P\otimes 1\otimes 1
,\\ & &
-\tau\Lambda^{-1}\otimes 1\otimes 1+
\Lambda^{-1}\otimes \tau\otimes 1 +
\Lambda^{-1}\otimes \Lambda\otimes \tau
\ket_{13})\ =\\
&=&\Sigma(P\tau-\gamma\tau-2 P)\ =\\&=&
\tau P -\gamma\tau-2P
\end{eqnarray*}
and similary the others.
\endproof

To obtain the classical limit one may introduce the new
generators
\bea
p&\defe&{P \over \alpha}\nn\\
t&\defe&{T \over \alpha}
\eea
where $\alpha \ \defe \ \gamma/2$
In these generators the quantum double $\DD(\CA)$ looks
\begin{eqnarray}
{[p,\ \tau]}&=& -2\tau-2 p\nn\\
{[p,\ \pi]}&=& \Lambda^{-1}e^{2\alpha t} -1+2\pi\nn\\
{[t,\ \tau]}&=& {1\over \alpha}(\Lambda-1)\nn\\
{[t,\ \pi]}&=& 0\nn\\
{[\tau,\ \pi]}&=& -2\pi\nn\\
{[t,\  p]}  &=&-{1\over \alpha} ( e^{2\alpha t}-1)\nn\\
\Delta (\tau)&=& \tau\otimes 1 + \Lambda\otimes\tau\nn\\
\Delta (\pi)&=& \pi \otimes 1+ \Lambda^{-1}\otimes\pi\nn\\
\Delta(t) &=& t\otimes 1 + 1\otimes t\nn\\
\Delta(P) &=& P\otimes 1+e^{2\alpha t}\otimes p
\end{eqnarray}

The classical limit is then the enveloping algebra of
the Lie algebra $L$ with four generators and Lie brackets
have this form:
\begin{eqnarray}
{[\tau,\ \pi]}&=& -2\pi\nn\\
{[p,\ \tau]}&=& -2\tau-2p\nn\\
{[t,\ p]}&=&- 2t\nn\\
{[t,\ \tau]}&=& 2\pi\nn\\
{[p,\ \pi]}&=& 2t\nn\\
{[t,\ \pi]}&=& 0
\end{eqnarray}

\section{The universal \protect $R$-matrix}
We have to find the dual bases for $\CA$ and
$\CA^o$. If define $R_j$ as
\bea
R_1&\defe &  P\nn\\
R_n&\defe & P^n - \sum_{k=1}^{n-1}\Theta^n_k P^k (-\gamma)^{n-k}
\eea
we have
\be
\bra T^n R_k,\ \tau^m \pi^l\ket \ = \ n!k! \delta_{nm}\delta_{kl}
\ee
\proof
easy
\endproof

Then we have the universal $R$-matrix in the form
\be
\CR \ =\ \sum_{n,k=0}^{\infty} {1 \over n! k!}
\tau^n \pi^k\otimes T^n R_k
\ee

\section{Representations of \DD(\CA)}
\vspace{10pt}

The problem to find (all (irreducible)) representations
of \DD(\CA) is generally very dificult, but in our
case we have luckily at least two representations
obtained from the classical limit $L$:

The two dimensional one $\rho$, parametrized by the
three complex numbers $x,\ y,\ z$
\bea
\rho_{xyz}(\tau)&\defe& -\rho_{xyz}(p)\
\defe \ \sigma_3 + x \one + y \sigma_-\\
\rho_{xyz}(\pi)&\defe& \rho_{xyz}(t)\
\defe \ z \sigma_-
\eea
where $\one$ is the identity matrix and $\sigma_3,\
\sigma_-$ are Pauli matrices
\be
\sigma_3=\pmatrix{ 1&0\cr
                   0&-1}\quad\quad
\sigma_-=\pmatrix{ 0&0\cr
                   1&0}
\ee
Because $ \sigma_-^2=0$ the representation $\rho$ of $L$
induce the representation of \DD(\CA) and so we have this
solution of YBE:
\be
R_{x_1,y_1,z_1,x_2,y_2,z_2}
= \pmatrix{
     1& 0& 0& 0 \cr
   - \alpha z_1 (1+x_2)& 1& 0& 0 \cr
  \alpha z_2 (1+x_1) & 0& 1& 0 \cr
   f
 &\alpha  z_2(x_1-1)
 &-\alpha z_1(x_2-1)& 1 }
\ee
where $f=-\alpha^2 z_1 z_2 (x_1-1)( x_2+1)+\alpha (z_2 y_1-z_1 y_2)$.
The solution of constant Yang-Baxter equation (setting
$x_1=x_2=x,\ y_1=y_2=y,\ z_1=z_2=z$ and $ \gamma=\alpha z(x+1),\
\beta =\alpha z (x-1) $)
\be
R_{\beta\gamma} = \pmatrix{
     1& 0& 0& 0 \cr
   - \gamma& 1& 0& 0 \cr
  \gamma & 0& 1& 0 \cr
   -\gamma\beta &\beta
 &-\beta&1 }
\ee
 is the transposed~(\ref{man}).

The three dimensional representation we may obtain using the fact
that the element $\pi-t$ is central in $L$ and after the
 factorization $L/(\pi-t)$ we have the 3D Lie algebra $\tilde{L}$
with the structure
\begin{eqnarray}
{[\tilde{\tau},\ \tilde{\pi}]}&=& -2\tilde{\pi}\nn\\
{[\tilde{p},\ \tilde{\tau}]}&=& -2\tilde{\tau}-2\tilde{p}\nn\\
{[\tilde{\pi},\ \tilde{p}]}&=&- 2\tilde{\pi}
\label{tl}
\eea
The representation $\varrho$ is given by the adjoint representation
\bea
 \ad\tilde{ \pi}&=&  \pmatrix{0&2&-2\cr
                        0&0&0\cr
                        0&0&0}\nn\\
\ad\tilde{ \tau}&=&  \pmatrix{-2&0&0\cr
                        0&0&2\cr
                        0&0&2}\nn\\
\ad\tilde{ p}&=&      \pmatrix{2&0&0\cr
                        0&-2&0\cr
                        0&-2&0}\\
\eea
as follows
\bea
\varrho_{xy} (\pi)&=& \varrho_{xy} (t)\ =\ y\cdot\ad\tilde{\pi}\nn\\
\varrho_{xy} (\tau)&=& \ad\tilde{\tau} + x \one\nn\\
\varrho_{xy} (p)&=& \ad\tilde{p} - x \one.
\eea

As above since $\ad\tilde{ \pi}^2=0$ this representation
gives the solution of YBE
\bea
R(x_1,y_1,x_2,y_2)&=& \one + \alpha (\varrho_{x_1 y_1} (\pi)\otimes
\varrho_{x_2 y_2}(p)+ \varrho_{x_1 y_1} (\tau)\otimes
\varrho_{x_2 y_2} (t))+\nn\\ & &+
\alpha^2 \varrho_{x_1 y_1} (\tau\pi) \otimes \varrho_{x_2 y_2} (t p)
\eea

\section{Conclusion and Remarks}
\vspace{10pt}

The classical limit of the quantum double \DD(\CA) -
Lie algebra $L$ is obviously not $gl(2)$ as one may see
after the change of bases in $\tilde{L}$ (\ref{tl}),
which provides such commutation relations:
\be
{[H,\ X^{\pm}]}\ =\ \pm 2 X^{\pm}\quad [X^{+},\ X^{-}]\ =\ 0
\ee

After the finishing of this letter we have obtained the
preprint \cite{Vla:ha93}, where the author reformulates
the FRT method in terms of the Quantum double. He has
studied the case of the matrix (\ref{man}) and
he has constructed the corresponding Quantum double (but
the formula for the universal $R$-matrix is given only
in several terms of its formal power expansion).
It is however generally very dificult to compare those
explicit formulae for two algebras.
 We have solved this problem only for the classical limit
and we are able to show the equivalence.

\end{document}